\documentstyle[12pt,aps]{revtex}
\begin{document}
\title{Signature inversion in axially deformed $^{160,162}$Tm}
\author{ J.~Kvasil$^1$, R.G.~Nazmitdinov$^{2,3}$, A.~Tsvetkov$^1$,
and P.~Alexa$^4$}
\address{$^1$ Institute of Particle and Nuclear Physics, Charles University,
V Hole\v{s}ovi\v{c}k\'ach 2, CZ-180 00 Prague 8, Czech Republic\\
$^2$ Max-Planck-Institut f\"ur Physik komplexer
Systeme, D-01187 Dresden, Germany\\
$^3$Bogoliubov Laboratory of Theoretical Physics,
Joint Institute for Nuclear Research, 141980 Dubna, Russia\\
$^4$ Institute of Chemical Technology, Technick\'a 5,
CZ-166 28 Prague 6, Czech Republic}
\maketitle

\vskip 1 cm

\begin{abstract}
{\sf The microscopic analysis of experimental data in $^{160,162}$Tm
is presented within the two-quasiparticle-phonon model.
The model includes the interaction between odd quasiparticles
and their coupling with core vibrations. The coupling explains 
naturally the attenuation of the Coriolis interaction in rotating 
odd-odd nuclei. It is shown that the competition between the 
Coriolis and neutron-proton interactions is responsible
for the signature inversion phenomenon.}
\end{abstract}

\vskip 0.3cm

\centerline{PACS number(s): 21.10.Re,21.60.Ev,21.60.Jz,27.70.+q}

\vskip 0.3cm

Progress in the development of large arrays of $\gamma$--detectors
led to observations of various phenomena in rotating
nuclei. In many cases rotational states can be characterised
by the signature quantum number $r=e^{-i\alpha\pi}$ which
defines the admissible spin sequence for a rotational band
according to the relation $I=\alpha+2n (n=0,1...)$.
It is associated with the $D_2$ spatial symmetry
of non-rotational degrees of freedom of a nuclear system \cite{BM}. 
In particular, in experimental data we observe two signature
partner bands with $r=+1(\alpha=0)$ and $r=-1(\alpha=1)$ 
in even-mass nuclei and
$r=\pm 1(\alpha=\pm \frac{1}{2})$ in odd-mass nuclei
($\alpha=-\frac{1}{2} \equiv \frac{3}{2}$), which are
separated by a signature splitting energy.
The band, which is lower in energy, is called the favored band.
It is expected that the signature splitting should increase with
a growing of the angular momentum I. 
However, for some nuclei the signature splitting
decreases with spin and, furthermore,
the unfavoured band becomes lower in energy than
the favoured one. This phenomenon is
called the signature inversion and attracts the
experimental and theoretical attention, since it is not
completely understood. 

For example, the low-spin signature inversion of the
$\pi (h_{11/2})\otimes\nu (i_{13/2})$ band has been studied
systematically in odd-odd nuclei of the $A\sim 160$
region \cite{1}. The analysis within the Cranked Shell Model (CSM)
\cite{Ben} suggests that the signature inversion 
could be only observed in the region of $62<Z<70$ and
the phenomenon is a consequence of the triaxiality.
However, these predictions are not consistent with subsequent
observations of the low-spin signature inversion in nuclei
with $Z=71, 73$ \cite{1}. The signature inversion can be seen also
in bands different from $\pi(h_{11/2})\otimes \nu(i_{13/2})$.
Since the density of neutron-proton (n-p) two-quasiparticle
configurations is large in the low-lying part of the spectrum, 
one expects their strong mixing via residual
interactions of different nature.
It was proposed that the competition
between the Coriolis interaction (CI) and the n-p interaction
of an odd neutron and an odd proton may be responsible
for the signature inversion \cite{Rag,Rid}.
On the other hand, the coupling of odd quasiparticles with
vibrational excitations of an even--even core could be 
important as well.
Vibrational admixtures are essential elements 
for the description of energy spectra and transition probabilities 
in odd-odd nuclei and  
it was demonstrated in numerous calculations  
in the microscopic quasiparticle-plus-phonon model \cite{Sol,rev}.
We suggest that the interplay between the coupling of external nucleons
with core vibrations, the n--p interaction between external nucleons,
and the Coriolis interaction create an important mechanism of the
signature inversion phenomenon.
To this aim, we use the microscopic
two-quasiparticle+phonon+rotor model.
To distinguish between the considered mechanism 
and the one created by the triaxiality, 
the analysis has been done for two isotopes $^{160,162}Tm$.
Using the self-consistent cranking Hartee-Fock approach with 
the Skyrme III interaction (HF+Skyrme) \cite{Dob}, we obtained 
 very small equilibrium $\gamma$ deformations 
$(\gamma \sim 4^{0})$  for these nuclei
in the considered region of the angular momentum
$I\leq 28 \hbar$. In addition, 
the equilibrium axial  quadrupole deformation is quite stable: 
the quadrupole momentum $Q_{20}$ changes
from 1314 (1399) $fm^2$ to 1268 (1375) $fm^2$ in
$^{160}Tm(^{162}Tm)$ in the range of values of the rotational frequency 
0.1-0.25 MeV.

The low-lying states in odd-odd deformed nuclei can be described
within the adiabatic approximation of a separation of intrinsic,
non-rotational degrees of freedom and rotational ones, i.e., 
with the Hamiltonian $H=H_{rot}+H_{intr}$. 
These states were extensively investigated
theoretically for a long time \cite{rev}, however,
the main attention was paid to the Gallagher-Moszkowski
(GM) splitting \cite{18} and the Newby shift \cite{19}.
Since the $\gamma$-deformation is negligible for considered nuclei,
we use the Hamiltonian of the axially symmetric rotor model \cite{BM} 
\begin{equation}
\label{rot}
H_{rot}\!=\!\frac{\hbar ^2}{2{\it J}}\Big[
(\hat{I}^2-\hat{I}_3^2)-(\hat{I}_{+}\hat{j}_{-}+\hat{I}_{-}\hat{j}_{+})+
\frac{1}{2}(\hat{j}_{+}\hat{j}_{-}+\hat{j}_{-}\hat{j}_{+})\Big]
\end{equation}
where the first term  is a pure rotational term,
the second term represents the Coriolis interaction and the last
one is the centrifugal interaction.
Here ${\it J}$ is the moment of inertia of an odd-odd nucleus,
$\hat{I}_3$ is the projection of the angular momentum $(\vec{I})$ 
on the symmetry axis,
$\hat{I_{\pm}}=\hat{I_1}\pm i\hat{I_2}$, and 
$\hat{j_{\pm}}=\hat{j_1}\pm i\hat{j_2}$. The intrinsic angular momentum
$\vec{j}=\vec{j_n}+\vec{j_p}$ is a vector coupling of  
single-particle angular momenta of an odd neutron and odd proton.
Angular momentum is a good quantum number and the model has 
the advantage in comparison with the CSM  at low spin region for 
deformed nuclei.
The intrinsic part $H_{intr}$ consists of an axially 
deformed mean field $H_{sp}$,
a short-range residual interaction $H_{pair}$ (a monopole pairing),
the n-p interaction $H_{np}$ between an odd neutron
and an odd proton,
and of a long-range residual interaction, $H_{res}$, taken in the
form of the iso-scalar and iso-vector multipole decomposition
\begin{equation}
\label{resin}
H_{res}=-1/2\sum_{\lambda \mu \mu \geq 0}\sum_{\tau \tau^{'}}
(\kappa^{(\lambda \mu)}_0+\tau \tau ^{'}\kappa _1^{(\lambda\mu)})
Q^{(\tau )}_{\lambda \mu}Q^{(\tau ^{'} )}_{\lambda-\mu} \ .
\end{equation}

Here $Q^{(\tau )}_{\lambda \mu }$ is a symmetrised multipole
operator with a multipolarity $\lambda$ and projection $\mu$.
The index $\tau = -1$ and $+1$ corresponds to
neutron and proton systems, respectively.
Using the Bogoliubov transformation from single-particle
operators $(a_{\nu},a_{\nu}^{\dagger})$ to quasiparticle ones
$(\alpha_{\nu},\alpha_{\nu}^{\dagger})$ and the
random phase approximation (RPA), the
intrinsic Hamiltonian $H_{intr}$ can be represented in the
form $H_{intr}=H_{core}+H_{nO}+H_{pO}+H_{np}$.
The term $H_{core}$ generates quasiparticle and phonon (vibrational)
excitations of an even--even core, $H_{nO}$($H_{pO}$) describes
the coupling of odd neutron (proton) quasiparticles
with the core vibrations.
Explicit expressions for all terms involved in
$H_{intr}$ are given in \cite{rev}.
The eigenvalue problem  of the full Hamiltonian can be
solved in the basis of the symmetrised wave functions
$|I^{\pi}MK\varrho \rangle \sim
({\it D}^I_{MK} + (-1)^I{\it D}^I_{M-K}\hat{R_1})
|\psi_{\varrho }(K^{\pi})
\rangle$  \cite{BM}; $\varrho$ is the additional quantum number characterising 
the intrinsic state. The intrinsic 
wave function $|\psi_{\varrho }(K^{\pi})
\rangle$ corresponds to
the intrinsic energy $\eta_{\varrho K}$, i.e.,
$H_{intr}|\psi_{\varrho}(K^{\pi}) \rangle =
\eta_{\varrho K}|\psi_{\varrho}(K^{\pi}) \rangle $.
For $K^{\pi}=0^{\pm}$ the 
function  $|I^{\pi}MK\varrho \rangle$  is the eigenvector of the 
signature operator
${\hat R}_1=exp(-i\pi {\hat J}_1)$.
Consequently, the rotational band
with $K^{\pi}=0^{\pm}$ splits into the band with positive signature
states ($\alpha =0$) and even values of $I$ and the band
with negative signature states ($\alpha =1$) and
odd values of $I$.
First, we solve the RPA equations 
to determine structure and energies of
two-quasiparticle phonons $O^{\dagger}_{\lambda \mu}$ describing
the low-lying vibrational states of an even-even core.
Second, we solve the variational problem for the intrinsic
Hamiltonian $H_{intr}$. As a result, we obtain
the amplitudes $C^{\varrho}_{\nu _n\nu _p}$ of neutron-proton
two-quasiparticle components and the amplitudes
$D^{\varrho}_{\lambda \mu \nu _n\nu_p}$ of the coupling of
two-quasiparticle components with core vibrations
in the intrinsic wave function
$|\psi_{\varrho}(K^{\pi})\rangle$:

\begin{equation}
\label{wfint}
|\psi_{\varrho}(K^{\pi}) \rangle =
\Big( \sum_{\nu _n\nu _p}C^{K\varrho}_{\nu _n\nu _p}
\alpha^{\dagger}_{\nu _n}\alpha^{\dagger}_{\nu _p}+\sum_{\nu _n\nu _p
\lambda \mu}D^{\varrho}_{\lambda \mu \nu _n\nu _p}
\alpha^{\dagger}_{\nu _n}\alpha^{\dagger}_{\nu _p}
O_{\lambda \mu}^{\dagger}\Big)| \rangle
\end{equation}
Finally, we diagonalise the full Hamiltonian $H$ in which the intrinsic and 
rotational terms are coupled by the Coriolis interaction.

We remind the reader that in odd--odd nuclei one 
of the two-quasiparticle components
$\alpha _{\varrho _n}^{\dagger} \alpha _{\varrho _p}^{\dagger} | \rangle $
with the corresponding quantum number
$K=|K_{\varrho _n}\pm K_{\varrho _p}|$ dominates
in low-lying intrinsic states
$|\psi_{\varrho }(K) \rangle $ \cite{rev}. Two intrinsic states
with
$K_1=K_{\varrho _n}+ K_{\varrho _p}$ and
$K_2=|K_{\varrho _n}-K_{\varrho _p}|$ are
similar in structure (which means that amplitudes
$C^{\varrho K}_{\nu_n \nu _p}$ and
$D^{\varrho K}_{\lambda \mu \nu _n \nu _p }$ are similar) and they
form the well-known GM doublet with the corresponding GM splitting energy
$\Delta E^{(GM)}_{\varrho =\varrho _n\varrho _p}=
\eta _{\varrho K=K_{\varrho _n}+K_{\varrho _p}}-
\eta _{\varrho K=|K_{\varrho _n}-K_{\varrho _p}|}$.
Moreover, for the case of $K^{\pi}=0^{\pm}$, one can define the Newby shift
$\Delta E^{(N)}_{\varrho K=0}= \eta _{\varrho K=0\alpha =1}
-\eta _{\varrho K=0\alpha = 0}$
that determines the energy shift between two rotational
bands with the same internal structure but with different
quantum numbers $\alpha$. In other words, the Newby shift is 
the signature splitting energy for the signature partners with 
$K^{\pi}=0^{\pm}$ at the beginning of the signature splitting.

As discussed above, we estimated the equilibrium deformation in the
HF+Skyrme approach. To solve the RPA equations with density dependent
forces for rotating nuclei, especially for the odd-odd system, 
is quite difficult. This problem is still in its infancy and it needs 
a dedicated study. 
To carry out the numerical analysis for $^{160,162}$Tm,
the single--particle mean field $H_{sp}$ is
approximated by the Nilsson Hamiltonian with the parameters 
taken from \cite{Shel}. All shells up to $N=7$ are included
for neutrons and protons, respectively.
The deformation parameters which are similar to the HF+Skyrme estimations, 
the neutron and proton pairing gaps
for ground states obtained with the use of the Strutinsky method 
are taken from \cite{MN}.
According to the experimental systematics \cite{lbl}, 
in the vicinity of the
proton and neutron Fermi levels for the nuclei with $Z=69,71 (N=90,92)$ and
with $N=91,93 (Z=68,70)$ there are the following 
sequences  of single-particle states increasing with energy:
$$
protons:...3/2[411], 7/2[523], 1/2[411], 7/2[404], 1/2[541], 5/2[402],
9/2[514],...$$
$$
neutrons:...1/2[660], 1/2[400], 1/2[530], 3/2[532], 3/2[402], 3/2[651],
3/2[521],$$
$$ \qquad 5/2[642], 5/2523], 11/2[505],...$$
We reproduce these sequences and
the mean field Fermi levels for the $^{160,162}$Tm
are $1/2[411]$ and $3/2[521]$ for protons and neutrons, respectively. 
The BCS approximation is used to fix the number
of protons and neutrons.
The quadrupole and octupole multipoles have been taken as a residual
long-range interaction of the core. The corresponding
strength constants $\kappa^{2 \mu}, \kappa^{3 \mu}$ are
fitted in order to reproduce experimental
quadrupole and octupole one-phonon energies of the 
$^{158}Er$ (the core for $^{160}Tm$) and $^{160}Er$ (the core for 
$^{162}Tm$).
 The residual n-p interaction $H_{np}$
is taken in the form of $\delta$-force, i.e.
$H_{np} = \delta(\vec r _{p} - \vec r _{n})
(u_{0} + u_{1} \vec \sigma _{p} \vec \sigma _{p})$.
The parameters $u_{0}= -3.504$ MeV and $u_{1}= -0.876$ MeV
have been chosen from the systematics of
the GM splitting and Newby shifts for rare-earth
nuclei \cite{rev,22}.
All solutions of the RPA equations  up to
1 MeV have been taken into account at the diagonalisation procedure of
the full Hamiltonian $H$ for each angular momentum
and parity quantum numbers.

The signature inversion in both nuclei is observed 
in the negative parity bands.
To describe the experimental data in $^{160}$Tm
we included six negative parity bands in the diagonalisation procedure.
The inertial parameters 
$\hbar^2/2 \it J$ (see Eq.(\ref{rot})) of each band have been considered as 
variational parameters. The optimal
inertial parameter value is 11.2 keV 
for the ground rotational 
band and we used $\hbar^2/2 \it J\sim$ 11 keV 
for all other bands.
The calculations reproduce surprisingly well
the experimental signature inversion at $I\sim 18 \hbar$ 
(see Fig.1a) in the negative parity band. 
This band 
starts at $I=8\hbar$ as a band built on the intrinsic state with the largest 
p-n two-quasiparticle state 
$K^{\pi}=0^-(\pi 1/2[411]\otimes\nu 1/2[530])$.
The structure of this intrinsic state is following:
$(\pi1/2[411]\otimes\nu1/2[530])(75\%)$, 
$({(\pi1/2[411]\otimes\nu3/2[532])\otimes Q^{+}_{22}})(17\%)$,
$({(\pi3/2[411]\otimes\nu1/2[530])\otimes Q^{+}_{22}})(4\%)$,
$({(\pi5/2[413]\otimes\nu1/2[521])\otimes Q^{+}_{22}})(3\%)$,
$({(\pi1/2[411]\otimes\nu5/2[532])\otimes Q^{+}_{22}})(1\%)$.
The position of the signature inversion point depends on the 
relative values of the Newby shift and the strength of
the CI between this band 
and its GM partner band built on the 
$K^{\pi}=1^{-}(\pi 1/2[411]\otimes\nu 1/2[530])$ state. 
If we are limited by the independent quasiparticle approach,
it is necessary to introduce the attenuation factor for
the Coriolis interaction to reproduce the experimental data 
in the vicinity of the inversion point.
Thanks to collective phonon components in the intrinsic wave functions, 
Eq.(\ref{wfint}), the strength 
of the Coriolis interaction is reducing upon $\sim 25\%$.
Consequently, the phonon components fix the strength of the CI
self-consistently. 
The exact position of the signature inversion point can be obtained 
by the fitting of the value of the Newby shift, $\Delta E^{(N)}$ . 
The calculated value $\Delta E^{(N)}=120$ keV is closed 
to the optimal value $\Delta E^{(N)}=135$ keV 
of the Newby shift for this band.

Six negative parity bands 
are involved in the CI mixing calculations for $^{162}$Tm. 
The optimal values for the inertial parameters are similar to 
the ones of the $^{162}$Tm.
The calculated energy difference $[E(I)-E(I-1)]/2I$ for 
negative parity yrast states 
is compared with the experimental data in Fig.1b. 
The low spin region of the negative 
parity yrast rotational band (up to the spin $I\sim 6 \hbar)$ is built 
on the intrinsic state with the largest two-quasiparticle component 
$K^{\pi}=1^{-} (\pi 1/2[411]\otimes\nu 3/2[521])$. In the region $I>6\hbar$ 
(up to the spin
$I\sim 28\hbar$ where we stopped our calculations) the yrast band 
is built on the $K^{\pi}=0^-(\pi 1/2[411]\otimes\nu 1/2[530])$ state. 
Again, the vibrational admixtures decrease naturally the strength of the
Coriolis interaction. The experimental signature inversion
point at $ I\sim 16 \hbar$ is well reproduced in our calculations. 
The calculated  Newby shift $\Delta E^{(N)}$ is
$\sim 100$ keV which is slightly larger than
the fitted  Newby shift $\Delta E^{(N)}= 88$ keV.
It seems that the adiabatic approximation and our configuration 
space constitute a reasonable approach even at high spins. 
However, it should be 
necessary to compare the results with estimations within the cranking 
Hartree-Fock-Bogoliubov+RPA approach to make a final conclusion.
Since the main question is the underlying mechanism of the signature 
inversion, we believe that the model reproduces well enough 
the important features of the phenomenon.

Let us discuss in detail the physical mechanism of
the signature inversion in both nuclei.  
The largest component, 
$K^{\pi}=0^{-} (\pi 1/2[411]\otimes\nu 1/2[530])$, of the favoured band
contains the proton quasiparticle state from the $d_{3/2}$ sub-shell
and the neutron quasiparticle state
from the $h_{9/2}$ sub-shell. While the sign of the projections
 of the proton  $\vec s_{p}$ and neutron $\vec s_{n}$ 
spins onto the symmetry axes (z-axis) is the same, 
the sign of the z-projection of the proton 
intrinsic angular momentum $\vec j_{p}$
is opposite to the sign of the one of the
neutron intrinsic angular momentum $\vec j_{n}$.
The opposite situation holds for the corresponding GM partner
$K^{\pi}=1^{-} (\pi 1/2[411]\otimes \nu 1/2[530])$:
the sign of the z-projections of the neutron and proton angular 
momenta is the same, while the z-projections of the neutron and proton
spins have the opposite signs.
The configuration $K^{\pi}=0^{-} (\pi 1/2[411]\otimes\nu 1/2[530])$ 
 with the same sign of the z-projections of
the spins $\vec s_{n}$ and $\vec s_{p}$ (but with the opposite
signs of the z-projections of the
intrinsic angular momenta $\vec j_{n}$ and $\vec j_{p}$)
has the lowest excitation energy
due to the n-p interaction and this is consistent with
the empirical GM rule.
The CI has a tendency to
lower the energy of the n-p state where the intrinsic angular momenta
$\vec j_{n} = \vec l_{n} + \vec s_{n}$ and
$\vec j_{p} = \vec l_{p} + \vec s_{p}$ are aligned in parallel
($K^\pi=1^-$ state).
Therefore, when the lower partner of the GM splitting has
the same sign of the z-projections of the neutron $\vec s_{n}$
and proton $\vec s_{p}$ spins but the signs of the z-projections of 
the intrinsic angular
momenta $\vec j_{n}$ and $\vec j_{p}$ are opposite, the
competition between the n-p interaction and the CI could lead to the
crossing of the GM partners and, consequently,
to the signature inversion for a particular spin.

This mechanism can be illustrated in a simple
two-level model. The matrix of the full Hamiltonian 
for the GM doublet ($K^{\pi}=0^{-}$ and $K^{\pi}=1^{-}$ bands)
coupled by the CI (expressions for the matrix elements are given
in \cite{15}) can be written as

\begin{equation}
\label{matrix}
\pmatrix{A + \frac{\hbar ^2}{2{\it J}}
I(I+1) - (-1)^I a & \frac{\hbar ^2}{2{\it J}}[c \! - \! (-1)^I b]
\sqrt{I(I+1)} \cr
\frac{\hbar ^2}{2{\it J}} [c\! - \! (-1)^I b] \sqrt{I(I+1)} &
B + \frac{\hbar ^2}{2{\it J}} [I(I+1)-1]\cr }
\end{equation}

The term in the upper left corner of the matrix corresponds to
the unperturbed GM partner with the lowest energy ($K^{\pi}=0^{-}$
band). The term in the lower right corner corresponds
to the unperturbed GM partner with a highest energy ($K^{\pi}=1^{-}$
band). In Eq.(\ref{matrix}), we use the following notation:
$A= -\Delta E^{(GM)}/2$, $B=\varepsilon_1+\Delta E^{(GM)}/2$,
$\Delta E^{(GM)}$ is the GM splitting, $a$ includes a contribution
of the n-p interaction (a half of the Newby shift, $\Delta E^{(N)}/2$),
and the centrifugal interaction
which is smaller than $\Delta E^{(N)}/2$. The quantities
$|b|<|c|$ are generated by the
CI. The term $\varepsilon_1$ is the energy difference
between the contributions of the centrifugal interaction
in the non-perturbed $K^{\pi}=0^-$ and $K^{\pi}=1^-$
bands. For small $I$, the n-p interaction is dominant and
the CI can be neglected. Consequently, the matrix (\ref{matrix})
has diagonal non-zero matrix elements only,
and we obtain for yrast states
\begin{equation}
\label{1}
\Delta E(I)= E(I) - E(I-1) \approx
\Bigl\{ \begin{array}{ll} \frac{\hbar ^2}{{\it J}} I - 2 a &
\mbox{for}\ I \ \mbox{even,}\\
\frac{\hbar ^2}{{\it J}} I + 2 a & \mbox{for}\ I \ \mbox{odd.} \end{array}
\end{equation}
For large $I$, the CI is dominant, the n-p interaction and the
centrifugal interaction can be neglected, and we have
\begin{equation}
\label{2}
\Delta E(I)= E(I) - E(I-1) \approx
\Bigl\{ \begin{array}{ll} \frac{\hbar ^2}{{\it J}} I (1 + b) &
\mbox{for}\ I \ \mbox{even,}\\
\frac{\hbar ^2}{{\it J}} I (1 - b) & \mbox{for}\ I \ \mbox{odd.}
\end{array}
\end{equation}
for yrast states. From Eqs.(\ref{1}) and (\ref{2}) it follows that
if $a>0$, $b>0$, $(a<0$, $b<0)$, the yrast band consists of the states
with even (odd) values of the angular momentum in
the low-spin region,
while the odd (even) spin states form the yrast band in the high-spin region.
Signs of $a, b, c$ are determined by the signs of the intrinsic 
matrix elements 
of the n-p interaction, the Coriolis interaction, and 
the centrifugal interaction.
Consequently, the result depends on the structure of 
the intrinsic wave functions,
Eq.(\ref{wfint}), for both $K^{\pi}=0^{-}$ and $K^{\pi}=1^{-}$ 
bands.

In conclusion, 
the signature inversion in  negative parity bands of $^{160,162}$Tm is
described in the microscopic model which includes the coupling of
quasiparticle excitations with core vibrations.
Vibrational components of rotational states lead to the
attenuation of the Coriolis interaction which is crucial for
a correct description of the signature inversion point.
The competition between the Coriolis interaction
and the neutron-proton interaction  between odd quasiparticles
explains the mechanism of the signature inversion 
in $\pi(d_{3/2})\otimes \nu(d_{9/2})$ band of odd-odd axially deformed 
rotating nuclei. 
It is different, for example, 
from the signature inversion mechanism
in the negative parity band in $^{72}Br$, which is due to the onset of the
triaxiality \cite{Pl}.
These two mechanisms complement each other and, in principle,
should be included as main ingredients of the model of the signature
inversion.
The measurement of electromagnetic transitions could provide
a more detailed understanding of the contribution of phonon components in
the structure of excited states of odd-odd nuclei and their role in the
signature inversion phenomenon.

This work was partly supported by the Grant Agency of Czech Republic
(contract N 202/99/1718), the Russian Foundation for Basic Research
under Grant 00-02-17194 and the Votruba-Blokhintcev program of the BLTP of
JINR.

{\bf Figure Caption}\\

{\bf Fig.1} 
The energy difference 
 $(E(I)-E(I-1))/2I$ $(keV/\hbar)$ vs $I (\hbar)$ for the
negative parity yrast bands in the odd-odd nuclei: (a) $^{160}$Tm,
(b) $^{162}$Tm. The signature inversion
points are shown by arrows. 
Full triangles correspond to 
the result of calculations, 
empty triangles correspond to 
the experimental data from \cite{1,lbl}.
Lines connecting symbols are used to guide the eye.
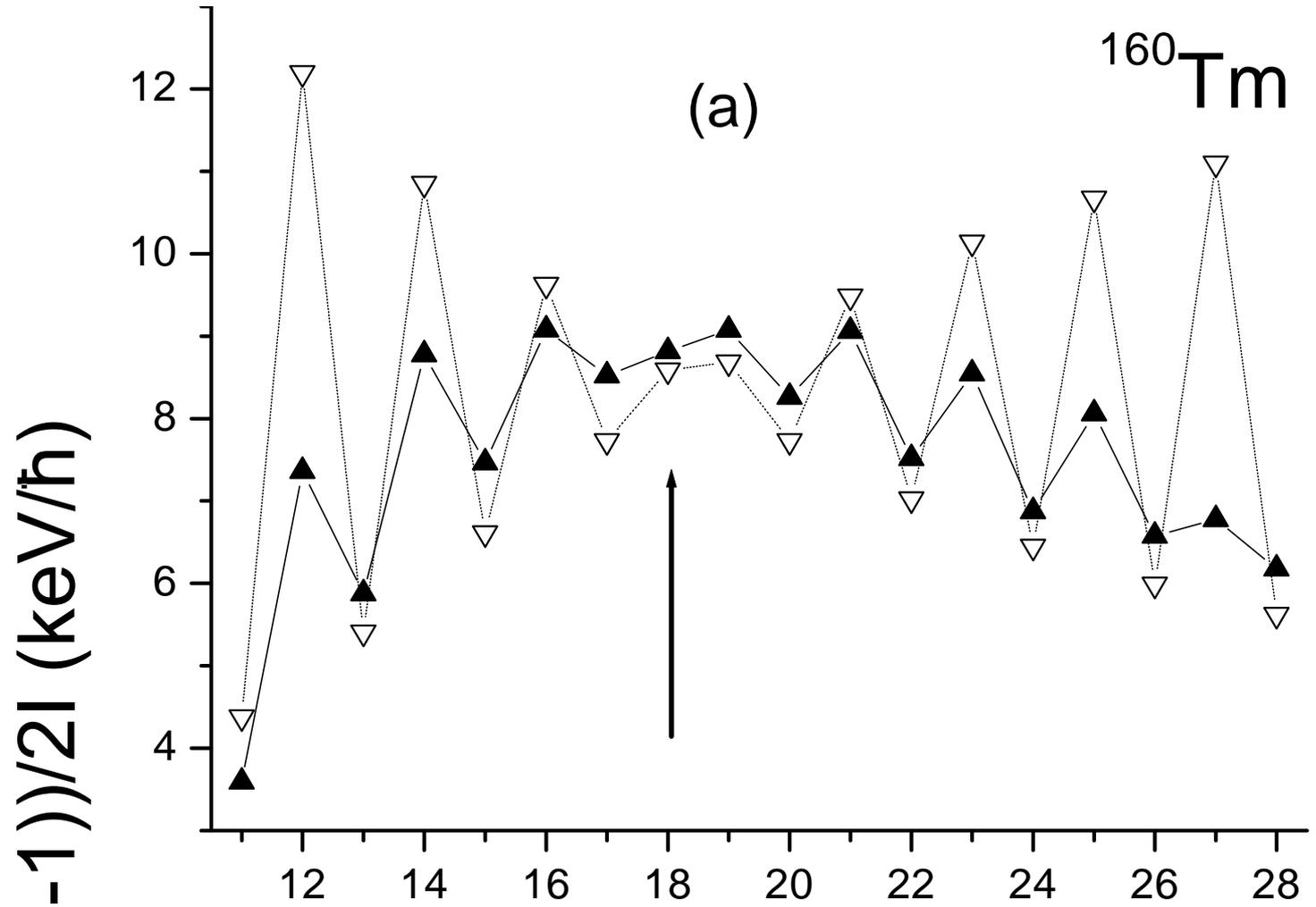
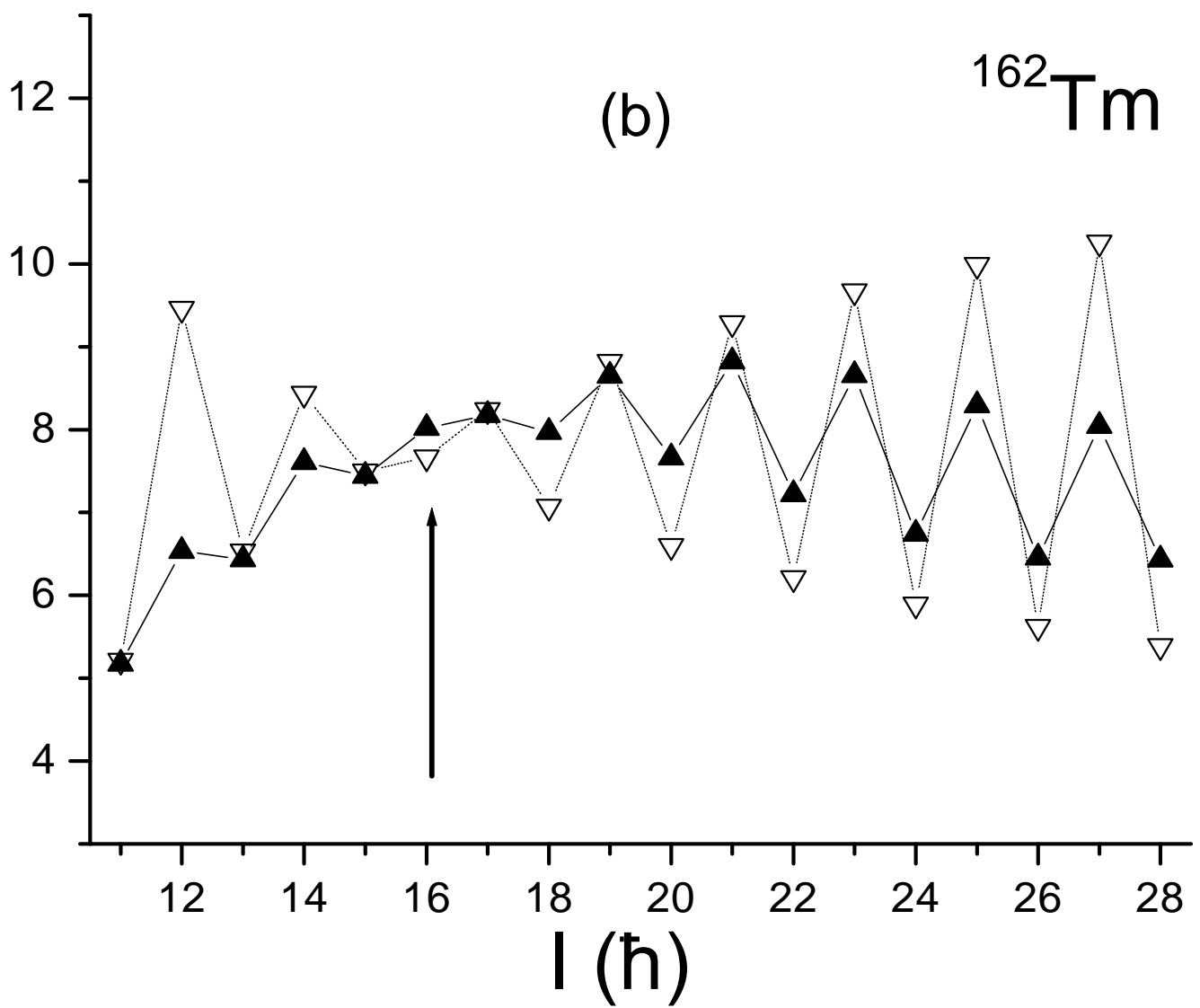
\end{document}